\documentclass[pre,aps,showpacs,twocolumn]{revtex4}
\usepackage{epsfig}
\usepackage{bm}

\begin{document}

\title{Gaussian paradox and clustering in intermittent turbulent signals}

\author{\small  A. Bershadskii}
\affiliation{\small {\it ICAR, P.O.\ Box 31155, Jerusalem 91000, Israel}}

\begin{abstract}
A relation between intermittency and clustering phenomena in 
velocity field has been revealed for homogeneous fluid turbulence. 
It is described how the intermittency exponent can be split into 
sum of two other exponents. One of these exponents (cluster-exponent) characterizes 
clustering of the 'zero'-crossing points in nearly Gaussian velocity field and another exponent 
is related to the tails of the velocity probability distribution. 
The cluster-exponent is uniquely determined by 
the energy spectrum of the nearly Gaussian velocity field and entire dependence 
of the intermittency exponent on Reynolds number is determined by the cluster-exponent. 

\end{abstract}

\pacs{47.27.-i, 47.27.Gs}

\maketitle

Relation between intermittency of dissipation field \cite{fs},\cite{sa},\cite{g} 
and clustering in turbulent velocity field \cite{sb1,sb2} is still very obscure. 
Taking into account that turbulent 
velocity itself is nearly Gaussian \cite{my} the clustering phenomenon in the velocity 
field should be also a 'Gaussian' one, while the intermittency phenomenon is usually associated 
just with non-Gaussian properties of the velocity derivatives \cite{fs},\cite{sa},\cite{my}. 
On the other hand, it is expected that high frequency events in the velocity field should 
provide significant contribution to the turbulent dissipation and, especially, 
to its high order moments \cite{fs},\cite{sms},\cite{bt}. 
\begin{figure} \vspace{-0.5cm}\centering
\epsfig{width=.45\textwidth,file=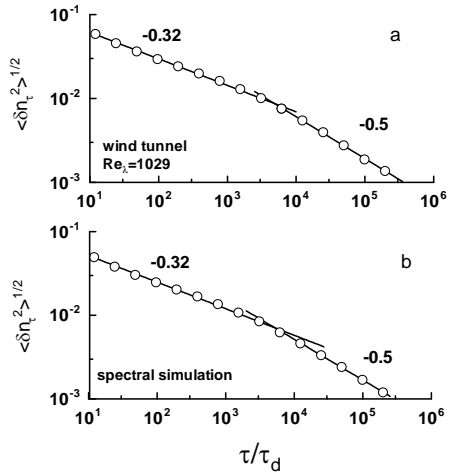} \vspace{-4cm}
\caption{Standard deviation of the running density fluctuations
against $\tau$ for for the velocity fluctuations measured
in the wind-tunnel \cite{pkw} (Fig. 1a) and for their Gaussian spectral simulation
(Fig. 1b).}
\end{figure}
\begin{figure} \vspace{-1cm}\centering
\epsfig{width=.45\textwidth,file=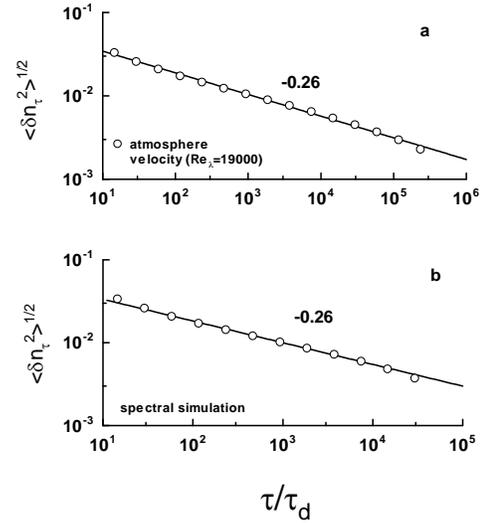} \vspace{-3.5cm}
\caption{As in Fig. 1 but for atmospheric surface layer at 
$Re_{\lambda}=19000$ \cite{sd} (Fig. 2a) and for its Gaussian spectral simulation (Fig. 2b).}
\end{figure}
Therefore, to find relationship between the two phenomena is rather a 
non-trivial task that demands special tools. In our recent papers \cite{sb1},\cite{sb2} 
we introduced a cluster-exponent to describe quantitatively clustering 
in turbulent velocity field and we found an empirical relationship between 
the cluster- and intermittency exponents. In present paper we will describe how the 
intermittency exponent can be split into sum of two other exponents 
(Eq. 6). One of these exponents is the cluster-exponent \cite{sb1} and another exponent 
is related to the tails of the velocity probability distribution. 
We also will show that the cluster-exponent is uniquely determined by 
the energy spectrum of the nearly Gaussian velocity field. This simple splitting 
of the intermittency exponent on the 
Gaussian and 'tail' components can be considered as 
a solution of the problem which was formulated above. Moreover, for finite Reynolds 
numbers $Re_{\lambda}$ the energy spectrum depends on $Re_{\lambda}$ and, therefore, 
the cluster-exponents depends on $Re_{\lambda}$, while the 'tail' component of the 
intermittency exponent is independent on $Re_{\lambda}$. Therefore, the entire dependence 
of the intermittency exponent on $Re_{\lambda}$ is uniquely determined by its Gaussian component 
(through the clustering phenomenon) that we will call Gaussian paradox.  \\     

Let us count the number of 'zero'-crossing points of the signal in a
time interval $\tau$ and consider their running density
$n_{\tau}$. Let us denote fluctuations of the running density as
$\delta n_{\tau} = n_{\tau} - \langle n_{\tau} \rangle$, where the
brackets mean the average over long times. We will be interested in
the scaling variation of the standard deviation of the running density
fluctuations $\langle \delta n_{\tau}^2 \rangle^{1/2}$ with
$\tau$ 
$$
\langle \delta n_{\tau}^2 \rangle^{1/2} \sim \tau^{-\alpha_n}
\eqno{(1)}
$$
For white noise it can be derived analytically \cite{molchan},\cite{lg} that 
$\alpha_n = 1/2$ (see also \cite{sb1}). In Figure 1a and 2a we show calculations of the standard 
deviation for a turbulent velocity signal obtained in the wind-tunnel experiment \cite{pkw} 
and for a velocity signal obtained in an atmospheric experiment \cite{sd} respectively.
The straight lines are drawn in the figures to indicate scaling
(1). One can see two scaling intervals in Fig. 1. The left
scaling interval covers both dissipative and inertial ranges,
while the right scaling interval covers scales larger then the
integral scale of the flow. While the right scaling interval is
rather trivial (with $\alpha_n =1/2$, i.e. without clustering), the
scaling in the left interval (with $\alpha_n < 1/2$) indicates
clustering of the high frequency fluctuations. The
cluster-exponent $\alpha_n$ decreases with increase of
$Re_{\lambda}$, that means increasing of the clustering (as it
was expected from qualitative observations). 

To describe intermittency a running average of fluctuations of the dissipation 
rate $\varepsilon (t)$ of the velocity signal $u(t)$ can be used \cite{po} 
(see also a discussion in \cite{lp})
$$
\langle \delta \varepsilon_{\tau}^2 \rangle^{1/2} \sim
\tau^{-\alpha_{\varepsilon}}  \eqno{(2)}
$$
where
$$
\varepsilon_{\tau} = \frac{\int_0^{\tau} \varepsilon (t) dt}{\tau} \eqno{(3)},
$$
and $\langle \delta \varepsilon_{\tau}^2 \rangle^{1/2}$ is standard deviation 
of $\varepsilon_{\tau}$.
Figure 3 shows an example of the scaling (2) providing the intermittency exponent 
$\alpha_{\varepsilon}$ for the wind-tunnel data (see \cite{po} for more details and Fig. 5 for 
the statistical errors).
\begin{figure} \vspace{-0.5cm}\centering
\epsfig{width=.45\textwidth,file=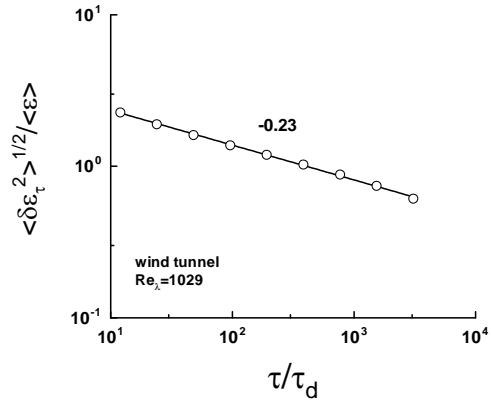} \vspace{-4.5cm}
\caption{Intermittency exponent as slope of the
straight line for the longitudinal velocity signal obtained in
the wind tunnel experiment \cite{pkw} at $Re_{\lambda} = 1029$.}
\end{figure}
\begin{figure} \vspace{-0.5cm}\centering
\epsfig{width=.45\textwidth,file=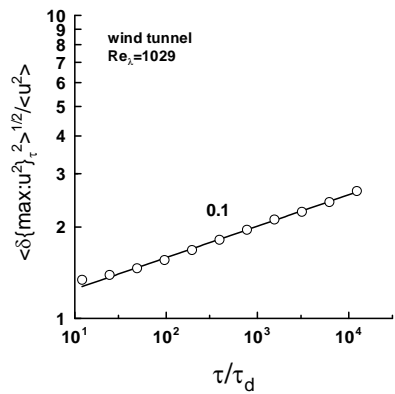} \vspace{-4cm}
\caption{ Statistical ('tail') exponent $\alpha_m \simeq 0.1$ as slope of the
straight line for the longitudinal velocity signal obtained in
the wind tunnel experiment \cite{pkw} at $Re_{\lambda} = 1029$.}
\end{figure}

Events with high concentration of the zero-crossing points in the intermittent 
turbulent velocity signal provide main contribution to the $\langle \delta \varepsilon_{\tau}^2 \rangle$ 
due to high concentration of the statistically significant local maximums 
in these events \cite{bt}. Therefore, one can make use 
of statistical version of the theorem 'about mean' in order to estimate the standard deviation
$$
\langle \delta \varepsilon_{\tau}^2 \rangle^{1/2} \sim  \langle 
\delta \{\max : u^2\}_{\tau}^2 \rangle^{1/2} \langle \delta n_{\tau}^2 \rangle^{1/2}  
\eqno{(4)}
$$
where $\{\max : u^2\}_{\tau}$ is maximum of $u(t)^2$ in 
the interval of length $\tau$. 
Value of $\langle \delta \{\max : u^2\}_{\tau}^2 \rangle^{1/2}$ should grow 
with $Re_{\lambda}$ (due to increasing of $\langle  u^2 \rangle$) and, 
statistically, with the length of interval 
$\tau$ (the later growth is expected to be self-similar), i.e.
$$
\langle \delta \{\max : u^2\}_{\tau}^2 \rangle^{1/2} = C(Re_{\lambda})
\cdot \tau^{\alpha_m} \eqno{(5)}
$$
where constant $C(Re_{\lambda})$ is a monotonically increasing function of $Re_{\lambda}$, 
and the statistical ('tail') exponent $\alpha_m \geq 0$ is independent on $Re_{\lambda}$. Fig. 4 
shows an example of scaling (5) providing the statistical ('tail') exponent 
$\alpha_m \simeq 0.10 \pm 0.01$.
Substituting (5) into (4) and taking into account (1) and (2) we can infer 
a relation between the scaling exponents
$$
\alpha_{\varepsilon}=\alpha_n-\alpha_m   \eqno{(6)}
$$
Unlike the exponent $\alpha_m$ the exponent $\alpha_n$ 
depends on $Re_{\lambda}$ (see Fig. 2 and Ref. \cite{sb1}). Therefore, we can learn from (6) 
that just the clustering
of the high frequency velocity fluctuations ($\alpha_n$ in (6)) is responsible for  
dependence of the intermittency exponent $\alpha_{\varepsilon}$ on the $Re_{\lambda}$.

It is naturally consider the exponent 
functions through $\ln R_{\lambda}$: $\alpha = \alpha (\ln Re_{\lambda})$
\cite{sb1},\cite{cgh},\cite{bg} (see also Appendix). Following to the general idea
of Ref. \cite{bg} (see also \cite{sb1}) let us expend this function for large
$Re_{\lambda}$ in a power series:
$$
\alpha (\ln Re_{\lambda}) = \alpha(\infty) + \frac{a_1}{\ln
Re_{\lambda}} + \frac{a_2}{(\ln Re_{\lambda})^2} + ... \eqno{(7)}
$$
 
\begin{figure} \vspace{-1.5cm}\centering
\epsfig{width=.45\textwidth,file=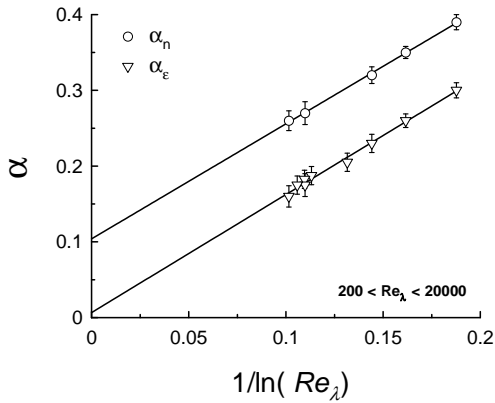} \vspace{-4.5cm}
\caption{The cluster-exponent: $\alpha_n$ (circles), against
$1/\ln Re_{\lambda}$ for velocity signal at different values of
Reynolds number ($200 < Re_{\lambda} < 20000$). Scaling exponent
$\alpha_{\varepsilon}$ (triangles) for fluctuations of the
dissipation rate of the velocity signals.}
\end{figure}
In Figure 5 we show the values of $\alpha_n$ (circles) 
and $\alpha _{\varepsilon}$ (triangles)
calculated for the velocity signals against $1/\ln Re_{\lambda}$ for $200 <
Re_{\lambda} < 20000$ (the data are taken from the wind-tunnel experiment \cite{pkw}, from the 
atmospheric experiment \cite{sd}, from a mixing layer and an atmospheric surface experiments 
\cite{po}). The straight lines (the best fit) indicate
approximation with the two first terms of the power series
expansion (7)
$$
\alpha_n (\ln Re_{\lambda}) \simeq 0.1 + \frac{3/2}{\ln
Re_{\lambda}},~~~~~~~~\alpha_{\varepsilon} (\ln Re_{\lambda}) \simeq \frac{3/2}{\ln
Re_{\lambda}} \eqno{(8)}
$$
This means, in particular, that
$$
\lim_{Re_{\lambda} \rightarrow \infty}\alpha_n \simeq 0.1,~~~~~ \lim_{Re_{\lambda} \rightarrow \infty}\alpha_{\varepsilon}\simeq 0 \eqno{(9)}
$$
The closeness of the constants $a_1 \simeq 3/2$ (the slopes of the straight lines in Fig. 5)
in the approximations (8) of the cluster-exponent $\alpha_n$ and of the intermittency exponent 
$\alpha_{\varepsilon}$ confirms the relationship (6) with $\alpha_m \simeq 0.1$ (cf Figs. 4 and 5). 

Thus one can see that entire dependence of the intermittency exponent 
$\alpha_{\varepsilon}$ on $Re_{\lambda}$ is uniquely determined by dependence 
of the cluster-exponent $\alpha_n$ on $Re_{\lambda}$. On the other hand, 
the cluster-exponent $\alpha_n$ itself is uniquely determined 
by energy spectrum of the velocity signal at suggestion that the velocity field is 
nearly Gaussian. This means that energy spectrum uniquely determines 
dependence of the intermittency exponent on the Reynolds number $Re_{\lambda}$. 
From conventional point of view on intermittency it seems as a paradox. 

We use spectral simulation to illustrate the uniquely relationship between energy spectrum and 
cluster-exponent for the Gaussian signals; i.e. we generate Gaussian stochastic signal with energy spectrum 
given as data (the spectral data is taken from the original velocity signals). Obviously, the 
pure Gaussian signal obtained in such simulation does not possesses the turbulent 
intermittency properties resulting in the estimate (4). 
Figures
1b and 2b show, as examples, results of such simulation for the wind-tunnel 
velocity data (Fig. 1b) and for the atmospheric surface layer (Fig. 2b).
Namely, Figures 1a and 2a show results of the calculations for the
original velocity signals and Figs. 1b, and 2b show corresponding
results obtained for the spectral simulations of the signals (to be sure, for the both 
Gaussian simulations $\alpha_{\varepsilon} \simeq 0.5$ unlike of $\alpha_n$). 

It is interesting 
to note that for isotropic turbulence with prominent inertial (Kolmogorov) range the 
only small-scale edge of the inertial range (this scale depends on $Re_{\lambda}$)
determines value of the cluster-exponent $\alpha_n$ corresponding to the spectrum. I.e. role of the 
viscous dissipation in this respect is reduced to the effective small-scale 'cut-off' effect and just 
dependence of the 'cut-off' scale on $Re_{\lambda}$ determines dependence of the 
intermittency exponent $\alpha_{\varepsilon}$ on $Re_{\lambda}$ through equation (6) 
for such flows. \\

I thank K.R. Sreenivasan for inspiring cooperation.
I also thank G. Falkovich, C.H. Gibson, A. Praskovsky, and V. Steinberg 
for discussions.

\section{Appendix}

Thin vortex tubes (or filaments) are the most prominent
hydrodynamical elements of turbulent flows \cite{batchelor}. 
The filaments are unstable in
3-dimensional space. In our recent paper \cite{sb1} we investigated this 
instability in order to renormalize the Taylor microscale Reynolds number. 

A straight line-vortex can readily develop a kink propagating along the filament with a constant speed. 
To estimate the velocity of propagation of such a kink 
let us first recall the properties of a ring vortex \cite{batchelor}. 
Its speed $v$ is related to its diameter $\lambda$ and
strength $\Gamma$ through
$$
v=\frac{\Gamma}{2\pi \lambda} \ln \left( \frac{\lambda}{2\eta}
\right),        \eqno{(A.1)}
$$
where $\eta$ is the radius of the core of the ring and
$\lambda/2\eta \gg 1$ (see figure 6).

If, for instance, a straight line-vortex develops a kink with a
radius of curvature $\lambda/2$, then self-induction generates a
velocity perpendicular to the plane of the kink. This velocity can
be also calculated using (A.1).
\begin{figure} \vspace{-2cm}\centering
\epsfig{width=.45\textwidth,file=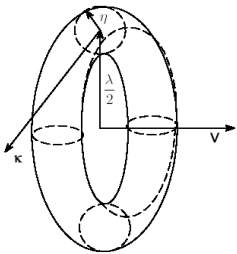} \vspace{-4.5 cm}
\caption{The vortex ring and its velocity.}
\end{figure}
One can guess that in a turbulent environment,
the most unstable mode of a vortex tube with a thin core of length
$L$ (integral scale) and radius $\eta$ (Kolmogorov or viscous scale), will be
of the order $\lambda$: Taylor-microscale \cite{my},\cite{saf}. Then, the
characteristic scale of velocity of the mode with the space scale
$\lambda$ can be estimated with help of equation (A.1). Noting
that the Taylor-microscale Reynolds number is defined as \cite{my}
$$
Re_{\lambda}=\frac{v_0 \lambda}{\nu},   \eqno{(A.2)}
$$
where $v_0$ is the root-mean-square value of a component of
velocity. It is clear that the velocity that is more relevant (at
least for the processes related to the vortex instabilities) for
the space scale $\lambda$ is not $v_0$ but $v$ given by (A.1).
Therefore, corresponding effective Reynolds number should be
obtained by the renormalization of the characteristic velocity in
(A.2), as 
$$
Re_{\lambda}^{eff}=\frac{v \lambda}{\nu} \sim \frac{\Gamma}{2\pi
\nu}\ln \left( \frac{\lambda}{2\eta} \right). \eqno{(A.3)}
$$

It can be readily shown from the definition that
$$
\frac{\lambda}{\eta} = const~ Re_{\lambda}^{1/2}  \eqno{(A.4)}
$$
where $const = 15^{1/4}\simeq 2$. Hence
$$
Re_{\lambda}^{eff}\sim \frac{\Gamma}{4\pi \nu}\ln (Re_{\lambda})
\eqno{(A.5)}
$$
The strength $\Gamma$ can be estimated as
$$
\Gamma \sim 2\pi v_{\eta}\eta  \eqno{(A.6)}
$$
where $v_{\eta}=\nu / \eta$ is the velocity scale for the
Kolmogorov (or viscous) space scale $\eta$ \cite{my}. 
Substituting (A.6) into (A.5) we obtain
$$
Re_{\lambda}^{eff}\sim \ln (Re_{\lambda}). \eqno{(A.7)}
$$
Thus, for turbulence processes determined by the vortex
instabilities the relevant dimensionless characteristic is $\ln
(Re_{\lambda})$ rather than $Re_{\lambda}$ (cf Eq. (7) and Fig. 5).

\end{document}